\documentclass[prl,twocolumn,amsfonts]{revtex4}
\usepackage{amsfonts}
\usepackage{amsmath}
\usepackage{bm}
\usepackage{epsf}

\newcommand{\beq}{\begin{equation}}
\newcommand{\eeq}{\end{equation}}
\newcommand{\beqa}{\begin{eqnarray}}
\newcommand{\eeqa}{\end{eqnarray}}

\newcommand{\ket}[1]{| #1    \rangle }
\newcommand{\bra}[1]{ \langle   #1  | }

\newcommand{\mel}[3]{  \langle #1  | #2   | #3  \rangle }
\newcommand{\amp }[2]{ \langle #1 |  #2  \rangle }

\newcommand{\weakv}[3]{\frac{ \mel{#1}{#2}{#3} }{ \amp{#1}{#3}} }

\newcommand{\vect}[1]{{\bm{ #1}}}

\newcommand{\xmin}{x_{\!{}_-}}
\newcommand{\pmin}{p_{\!{}_-}}
\newcommand{\xplus}{x_{\!{}_+}}
\newcommand{\pplus}{p_{\!{}_+}}

\newcommand{\hatxmin}{\hat{x}_{\!{}_-}}
\newcommand{\hatpmin}{\hat{p}_{\!{}_-}}
\newcommand{\hatxplus}{\hat{x}_{\!{}_+}}
\newcommand{\hatpplus}{\hat{p}_{\!{}_+}}
\newcommand{\hatxpm}{\hat{x}_{\!{}_\pm}}
\newcommand{\hatppm}{\hat{p}_{\!{}_\pm}}

\begin{document}

%%%%%%%%%%%%%%%%%%%%%%%%%%%%%%%%%%%%%%%%%%%%%%%%%%%%%%%%%%%%%%%%%%%%%%%%
%%%%%%%%%%%%%%%%%%%% Local Definitions %%%%%%%%%%%%%%%%%%%%%%%%%%%%%%%%%

\newcommand{\diracsl}[1]{\not\hspace{-3.0pt}#1}

\newcommand{\pin}{\psi_{\!{}_1}}
\newcommand{\psf}{\psi_{\!{}_2}}
\newcommand{\psv}[1]{\psi_{\!{}_#1}}
\newcommand{\lab}[1]{{}^{(#1)}}
\newcommand{\psub}[1]{{\cal P}_{{}_{\! \! #1}}}
\newcommand{\sst}[1]{{\scriptstyle #1}}
\newcommand{\ssst}[1]{{\scriptscriptstyle #1}}
\newcommand{\aft}{{{\scriptscriptstyle\succ}}}
\newcommand{\bef}{{{\scriptscriptstyle\prec}}}

\newcommand{\trans}[1]{\!{\xrightarrow{#1}}}

%%%%%%%%%%%%%%%%%%%%%%%%%%%%%%%%%%%%%%%%%%%%%%%%%%%%%%%%%%%%%%%%%%%%%%%%
%%%%%%%%%%%%%%%%%%%%%%%%%%%%%%%%%%%%%%%%%%%%%%%%%%%%%%%%%%%%%%%%%%%%%%%%

\title[Entanglement and precise inferences]
{Entanglement, weak values, and the precise inference of joint
measurement outcomes for  non-commuting observable pairs}

\author{  Alonso Botero }
\email{abotero@uniandes.edu.co} \affiliation{
    Departamento de F\'{\i}sica,
    Universidad de Los Andes,
    Apartado A\'ereo 4976,
    Bogot\'a, Colombia }\affiliation{ Department of Physics and Astronomy, University of
South Carolina, Columbia, SC 29208 }

\date{\today}

\begin{abstract}
\bigskip
The problem of inferring the outcome of a simultaneous measurement
of two non-commuting observables is addressed. We show that for
certain pairs  with dense spectra, precise inferences of the
measurement outcomes are possible in pre- and postselected
ensembles, and if the selections involve entangled states with
some other system. We show that the problem is related to the
problem of assigning weak values to a continuous family of
operators, and give explicit examples where this problem is
solvable. Some foundational implications are briefly discussed.

\end{abstract}

\maketitle

A joint measurement of two non-commuting observables
 can be understood as the simultaneous
coupling of the measured system to two independent instruments,
each of which is a good probe of either observable when coupled
individually~\cite{Busch95}. Independently of whatever meaning--if
any--one may choose to attach to  the instrument readings in this
context,
 no  state preparation can be realized so as to
systematically ensure a definite outcome  in any subsequent
measurement of this type~\cite{BuschLaht84}. Instead,  the mutual
back-action of the instruments ensures an uncertainty relation for
the  joint outcomes with a lower bound that is
  {\em twice} that given by the standard uncertainty principle~\cite{ArtGoo88}.

Fully accepting these facts, consider  a scenario in the same
spirit of the so called ``King's
problem"\cite{VAA87,Ben89,EngA01,Arav03,KimTanOz06}: Alice sends a
particle to Bob, which he then subjects to a joint measurement of
a non-commuting observable pair--say $\hat{x}$ and $\hat{p}\,$;
finally, he returns the particle to Alice. Suppose Bob chooses not
to reveal his readings. The question is: are there conditions
under which the readings can still be inferred, {\em precisely},
by Alice?

In this letter we show that this question can be answered in the
affirmative for certain operator pairs, and if she makes
appropriate measurements both before and after the particle has
passed through Bob's instruments. Moreover, she will need
entanglement resources in order to achieve the task, not unlike
other quantum inference problems where entanglement is required
for optimality (e.g., ~\cite{MassPop95,ChafBar97}).  Therefore, we
believe the results point to an interesting connection between
quantum uncertainty, non-locality and  ultimately, classicality.
These questions will be addressed briefly at the conclusion. For
the moment, we begin with the formal statement of the problem:

 {\em Joint
Measurement Inference Problem:} Let $(\hat{A}_1$, $\hat{A}_2)$ be
a pair of observables of a system  with no common eigenstates, and
$\hat{U}(q_1,q_2)$ a unitary operator of the form
\begin{equation}
 \hat{U}(q_1,q_2) = \exp[i (\hat{A}_1 {q}_1 + \hat{A}_2 {q}_2
 )]\,
\end{equation}
($\hbar\equiv 1$). When $(q_1,q_2)$ are operator-valued,
$\hat{U}(\hat{q}_1,\hat{q}_2)$  is the unitary evolution operator
describing an impulsive simultaneous interaction of the system
with two separate instruments $(I_1,I_2)$ described by canonical
 pairs $(\hat{q}_1, \hat{\pi}_1)$ and
$(\hat{q}_2,\hat{\pi}_2)$. Given this interaction, the problem is
to find a realizable set of conditions  not involving the
instruments  such that for an arbitrary initial state $\hat{\rho}$
of $(I_1,I_2)$, the conditional post-interaction density matrix
$\hat{\rho}'$ of the instruments,  is related to $\hat{\rho}$
through a quantum operation with a unitary normalized Kraus
operator
\begin{equation}
\hat{\rho}' = \hat{F}\hat{\rho}\hat{F}^\dagger\, , \ \ \ \hat{F} =
\exp\left[i (\alpha_1 \hat{q}_1 + \alpha_2 \hat{q}_2
 ) \right]\, ,
\end{equation}
for some $\hat{\rho}$-independent pair of real numbers
$(\alpha_1,\alpha_2)$. If the problem has a solution, then the
 conditional probability distribution of the ``pointer variables"
$(\hat{\pi}_1,\hat{\pi}_2)$ satisfies $ {\cal
P}(\pi_1,\pi_2|\hat{\rho}') = {\cal
P}(\pi_1-\alpha_1,\pi_2-\alpha_2|\hat{\rho})\, $, and the
conditional outcomes  can  be ascertained to within the
uncertainties in $(\pi_1,\pi_2)$. Thus, the outcomes
$(\alpha_1,\alpha_2)$  can be unequivocally ascertained for any
single trial in the ``sharp" limit $(\Delta \pi_1 , \Delta \pi_2)
\rightarrow 0$.

We  discuss solutions to the inference problem involving complete
pre- and postselection measurements, in which case the Kraus
operator is $\hat{F} \propto
\bra{\psi_f}\hat{U}(\hat{q}_1,\hat{q}_2)\ket{\psi_i}$ with pure
initial and final states $\ket{\psi_i}$ and $\ket{\psi_f}$.
Recalling the definition of the weak value of a quantum mechanical
observable\cite{VAA88,AV90},
$A_w=\bra{\psi_f}\hat{A}\ket{\psi_i}/\amp{\psi_f}{\psi_i}$, the
inference problem is then equivalent to finding a pair
$(\ket{\psi_i},\ket{\psi_f})$ such that weak values $e^{i
(\alpha_1 {q}_1 + \alpha_2 {q}_2)}$ can be assigned to all
elements of the continuous set of operators $\{ \hat{U}(q_1,q_2)|
(q_1,q_2) \in \mathbb{R}^2\}$. Alice's task should then be to
perform
 pre- and postselection measurements, yielding initial and final
 states for which
 this assignment is possible.

Let us then consider the case of the canonical variable pair
$(\hat{x},\hat{p})$ for a particle in one dimension. A natural
first guess at a solution would seem to be a pre-and postselected
ensemble defined by initial and final eigenstates of $\hat{p}$ and
$\hat{x}$ respectively. However, applying the Baker-Hausdorff
lemma to the exponential in $\bra{p}e^{i (\hat{x} q_1 + \hat{p}
q_2)}\ket{x}$, we find the Kraus operator $ \hat{F} = \, e^{i x
\hat{q}_1 + i p \hat{q}_2 }\, \ e^{\frac{i}{2}\hat{q}_1 \hat{q}_2}
\, $, differing from the desired form by the additional  term
$e^{\frac{i}{2}\hat{q}_1 \hat{q}_2}$.  This term represents the
back-action  between the two measurements, as it generates the
canonical transformation $(\delta \hat{\pi}_1, \delta \hat{\pi}_2)
=\frac{1}{2}(\hat{q}_2,\hat{q}_1)$; in the absence of correlations
in the preparation of the instruments, this back-action term can
be shown to enforce the uncertainty relations in the final
outcomes $(\pi_1', \pi_2')$
\begin{equation}\label{pointuncert}
\Delta \pi_1' \Delta \pi_2' \geq   \Delta \pi_1 \Delta \pi_2 +
\frac{1}{16\Delta \pi_1 \Delta \pi_2} \geq 1/2\, .
\end{equation}
The conditional uncertainty bounds now  coincide with standard
uncertainty relations,  suggesting  that even with additional
information from  postselection, the standard limit is still
unbreachable due to inevitable back-action.

Note, however, that we have only looked at complete measurements
on the system. So suppose instead that the selections involve
measurements of general observables of the system and an ancillary
canonical system, with canonical variables
$(\hat{x}_a,\hat{p}_a)$. In particular, consider the conjugate
pairs
 $(\hatxpm,\hatppm)$  defined by
\begin{subequations}
\begin{eqnarray}\label{transeq}
\hatxplus = (\hat{x} + \hat{x}_a)/2 \, & \ \ \ & \hatpplus =
\hat{p}+
\hat{p}_a \, ,  \\
\hatxmin = \hat{x} - \hat{x}_a \, & \ \ \ & \hatpmin = (\hat{p}
-\hat{p}_a)/2 \, ,
\end{eqnarray}
\end{subequations}
so that $\hat{x}= \hatxplus + \hatxmin/2$ and $\hat{p}=
\hatpplus/2 + \hatpmin$. The linear combination $ \hat{x}\,q_1 +
\hat{p}\,q_2$ can then be expressed as  the sum of the two
commuting combinations: $ (\hatxmin q_1+ \hatpplus q_2)/2\, $ and
$ \hatxplus q_1+  \hatpmin q_2$, each involving commuting
operators; thus, we can perform the factorization
\begin{equation}
e^{i (\hat{x} q_1 + \hat{p} q_2)}=e^{\frac{i}{2} (\hatxmin q_1 +
\hatpplus q_2)}\, e^{i (\hatxplus q_1 + \hatpmin q_2)} \, .
\end{equation}
Taking  eigenstates of the two terms in the factorization, for
instance an initial eigenstate of ($\hatxmin,\hatpplus$) and a
final eigenstate of $(\hatxplus, \hatpmin)$, we obtain the unitary
normalized
 Kraus operator
\begin{equation}
\label{krausformcan}
 \hat{F}=\weakv{\xplus \pmin}{e^{i (\hat{x}
\hat{q}_1 + \hat{p} \hat{q}_2)}}{\xmin \pplus} = e^{i (x \hat{q}_1
+  p \hat{q}_2)}\, ,
\end{equation}
with $x=\xplus + { \xmin / 2}$ and $p = \pmin + {\pplus/ 2}$. With
the absence of the  back-action term, the
 ensemble with $\ket{\psi_f}=\ket{\xmin
\pplus}$ and $\ket{\psi_i} =\ket{\xplus \pmin}$   solves the
inference problem for $(\hat{x},\hat{p})$.

This solution ensemble reveals another surprising result when we
consider, with otherwise the same conditions, the joint
measurement of all four canonical variables of the system plus the
ancilla with
 four different instruments. The corresponding
  Kraus operator then involves the quantity $
\bra{\xplus \pmin}e^{i (\hat{x}\hat{q}_1 + \hat{p}\hat{q}_2 +
\hat{x}_a \hat{q}_3 + \hat{p}_a \hat{q}_4)}\ket{\xmin \pplus}$.
Disentangling the exponential, we obtain
\begin{equation}\label{Krausfour}
\hat{F} = e^{i ({x}\hat{q}_1 + {p}\hat{q}_2 + {x}_a \hat{q}_3 +
{p}_a \hat{q}_4) } e^{ \frac{i}{2}(\hat{q}_1\hat{q}_4 +
\hat{q}_2\hat{q}_3)}\, ,
\end{equation}
where $x_a = \xplus - \xmin/2$ and  $p_a = -(\pmin-\pplus/2)$. It
is therefore the crossed pairs of {\em commuting} observables
$(\hat{x},\hat{p}_a)$ and $(\hat{p},\hat{x}_a)$, that generate a
back-action term  and  hence  uncertainty relations as those of
Eq.~(\ref{pointuncert}) for initially uncorrelated instruments.
These results, as the results of our earlier failed attempt, are
consistent with what could be conjectured to be a new type of
uncertainty relation for a composite system with two canonical
degrees of freedom, allowing for any pre- and postselection
described by states of the composite system only. In the absence
of any information about correlations in the preparation of the
measuring instruments, it appears as if the uncertainties in any
possible inference of the readings
$(\pi'_{x_1},\pi'_{p_1},\pi'_{x_2},\pi'_{p_2})$ from a joint
measurement of all canonical variables  are {\em only} constrained
by the  relation
\begin{equation} \Delta \pi'_{x_1}\,  \Delta
\pi_{p_1}'\, \Delta \pi_{x_2}' \Delta \pi_{p_2} '\geq 1/4 \, .
\end{equation}
It is as if entanglement allowed us to arbitrarily deform an
uncertainty volume in the ``inference phase-space", provided the
total  volume of $1/4$ is preserved.

Prior to further speculation, we turn our attention to the
assignment of weak values associated with the general inference
problem.
 The assignment problem can be cast in
terms of a complex linear map
\begin{equation}
W: \hat{A}  \rightarrow {\rm Tr}_{s}(\hat{A} \hat{W})\, , \ \ \
\hat{W} = {\rm Tr}_a\left(
\frac{\ket{\psi_i}\bra{\psi_f}}{\amp{\psi_f}{\psi_i}} \right)\, ,
\end{equation}
where $\hat{W}$, henceforth termed the {\em weak value operator},
 is realized by
 some pair ($\ket{\psi_i}$ , $\ket{\psi_f}$) of
non-orthogonal states in the total Hilbert of the system and an
ancilla. The general question is then to determine  for  a given
set of pairs $( \hat{A}^{(i)},\alpha^{(i)})$, an $\hat{W}$ such
that $W:\hat{A}^{(i)} \rightarrow \alpha^{(i)},\ \forall i$.

Consider the finite-dimensional case first. For the  space of
linear operators acting on a Hilbert space ${\mathcal H}_s$ of
dimension $d$, we introduce a basis
$\hat{E}_{1},....,\hat{E}_{d^2}$, orthonormal with respect to the
standard hermitian inner product: i.e., $(\hat{E}_{i}|\hat{E}_j) =
\delta_{ij}$ where $(\hat{A}|\hat{B}) \equiv{\rm
Tr}(\hat{A}^\dagger \hat{B})$.  An arbitrary observable $\hat{A}$
of the system  can then be represented as a vector $\vect{a}\in
\mathbb{C}^{d^2}$ through $ \hat{A} = \sum_i a_i \hat{E}_{i}\equiv
\vect{a}\cdot\hat{\vect{E}}\, $ with $a_{i} =
(\hat{E}_{i}|\hat{A})$, and where  the  dot product is euclidean
($\vect{a}\cdot\vect{b} \equiv \sum_i a_i b_i$). We will make an
exception for the weak value operator $\hat{W}$, which we choose
to expand  in terms of the {\em hermitian conjugate basis}: $
\hat{W} = \vect{w}\cdot\hat{\vect{E}}^\dagger$, in which case
${w}_a$ is the weak value of $\hat{E}_a$, and the weak value of
$\hat{A}$ can be written as $ \vect{a}\cdot\vect{w}$.  We reserve
the notation $\vect{I}$ for the  vector representing the identity
operator, where $I_i = {\rm Tr}(\hat{E}_i)^*$.  One constraint on
$\vect{w}$ is then $\vect{I}\cdot\vect{w}=1$.

The question of whether $\vect{w}$ has other constraints in
$\mathbb{C}^{d^2}$ besides $\vect{I}\cdot\vect{w}=1$ brings us to
the significance of entanglement. Suppose that one could write
$\hat{W}$ in the form $ \hat{W} =
\frac{\ket{\chi_i}\bra{\chi_f}}{\amp{\chi_f}{\chi_i}} $ for
normalized $\ket{\chi_i}$ and $\ket{\chi_f}$ in the system Hilbert
space. It is then easy to see that $\hat{W}$ satisfies the
relations $ \hat{W}^2 = \hat{W}$, $\ \hat{W}\hat{W}^\dagger =
\frac{\ket{\chi_i}\bra{\chi_i}}{|\amp{\chi_f}{\chi_i}|^2}$, and
$\hat{W}^\dagger\hat{W} =
\frac{\ket{\chi_f}\bra{\chi_f}}{|\amp{\chi_f}{\chi_i}|^2}$. These
conditions lead to non-trivial constraints involving the
components of the weak vector $\vect{w}$, the simplest of which
are:
\begin{equation}
\gamma_{i j}^* {w}_i{w}_j=1 \, \ \ {\rm and} \ \
\vect{w}\cdot\vect{w}^* = |\amp{\chi_f}{\chi_i}|^{-2} \geq 1 \, ,
\end{equation}
where $\gamma_{ij} = {\rm Tr}(\hat{E}_i\hat{E}_j)$. On the other
hand, define two vectors in a Hilbert space ${\cal H}_s\times{\cal
H}_a$, with $\dim(H_a) \geq d$
\begin{equation}
\ket{\Phi} = \sum_{i=1}^{n}\ket{i}_s\ket{i}_a \, , \ \  {\rm and}\
\ \ket{\vect{z}}=  \vect{z}\cdot \hat{\vect{E}}^\dagger
\ket{\Phi}\, .
\end{equation}
Using these as initial and final states for a pre-and
postselection  and tracing over the ancilla, we obtain  a
realization of $\hat{W}$ with $\vect{w} = \vect{z}/
(\vect{z}\cdot\vect{I})$, where no constraints are required for
$\vect{z}$ other than $\vect{z}\cdot \vect{I} \neq 0$. Thus we see
that with entanglement, it is possible to realize any complex
vector $\vect{w}$  that solves the associated linear-algebraic
problem of satisfying $\alpha^{(i)} = \vect{a}^{(i)}\cdot
\vect{w}$ for a vector set $\{\vect{a}^{(i)}\}$, provided the
problem is solvable and consistent with $
\vect{I}\cdot\vect{w}=1$.

Turning then to the linear algebra problem associated with $W:
U(q_1,q_2)\rightarrow e^{i (\alpha_1 q_1 + \alpha_2 q_2)}$, we now
show that the problem has no solution if $\hat{B}_{\theta} \equiv
\hat{A}_1 \cos\theta +\hat{A}_2 \sin\theta$ has a discrete
spectrum with $\theta$ in any finite subinterval
 of  $[0,\pi]$, as is always the case in the
finite-dimensional case. First note  by uniqueness of the
exponential expansion, that we must  have $W: \hat{B}_\theta^k
\rightarrow \beta_\theta^k$ where $\beta_\theta= (\alpha_1 \cos
\theta + \alpha_2 \sin\theta)$ for  all integer powers $k$; on the
other hand, by the Cayley-Hamilton theorem, we also know that
$\hat{B}_\theta$ annihilates its characteristic polynomial
$p_\theta(z) = \det(z\openone - \hat{B}_\theta)$, and hence, by
linearity of weak values, we conclude that
$p_\theta(\beta_\theta)=0$ and hence that $\beta_\theta$ must be
one of the eigenvalues
 $b_\theta^{(i)}$ of $\hat{B}_\theta$. But setting
$\theta=0$ or $\pi/2$ we see that $(\alpha_1,\alpha_2)$ must be a
pair of eigenvalues $(a_1,a_2)$ of $(\hat{A}_1, \hat{A}_2)$. The
conjunction of conditions can be visualized  as the
 intersection of the curve $\beta_\theta=a_1 \cos \theta +
a_2 \sin \theta $ with all zeroes of  $p_\theta(z)$ plotted as a
function of $\theta$. If $\hat{B}_\theta$ has a discrete spectrum
in some interval, then either the intersection occurs at discrete
values of $\theta$, in which case the solution fails, or else
there must exist a root behaving like $b^{(i)}_\theta=a_1
\cos\theta + a_2 \sin\theta $ in that interval. But this can only
occur if $\hat{A}_1$ and $\hat{A}_2$ have a common eigenstate, in
contradiction with our stated assumptions. Consequently, solutions
to the inference problem can only be found in the infinite
dimensional case and only for
 operators $\hat{A}_1$ and $\hat{A}_2$  such that the
combination $\hat{B}_\theta$ has a dense spectrum in some band
around $a_1 \cos\theta + a_2 \sin\theta$ for some pair of
eigenvalues $(a_1,b_1)$ of $(\hat{A}_1, \hat{A}_2)$.

Approximate assignments of the form $e^{i  \alpha_1 q_1 + \alpha_2
q_2 + o(q^s)}$ for some power $s$ can nevertheless be constructed
in the finite-dimensional case. A generic form for the leading
correction is obtained when $\hat{B}_\theta$ has a minimal
polynomial $m_\theta(z)$ of degree $s$ (for all $\theta$), and
assuming the linear independence of all symmetrized operators
$\hat{S}_{l,m}$ generated by the expansion $(\hat{A}_1 t +
\hat{A}_2)^k= \sum_{l=0}^{k}{k! t^l\over l!(k-l)!}
\hat{S}_{l,k-l}$ for all $k<s$. Then one can assign $W: (\hat{A}_1
q_1 + \hat{A}_2 q_2)^k\rightarrow (\alpha_1 q_1 + \alpha_2 q_2)^k$
for all $k<s$ with arbitrary $(\alpha_1,\alpha_2)$. The  leading
order term in the exponent is then  found to be $ -i^{s}|q|^s
m_\theta(z)/s! $, evaluated at $z=(\alpha_1 \cos\theta + \alpha_2
\sin\theta)$ (An example is a pair of orthogonal spin directions,
in a spin $j$ representation, in which case $m_\theta(z) =
p_\theta(z)$ and $s=2j+1$).

Proceeding with the  continuous variable case for a
one-dimensional canonical system, we generalize previous results
for  two standard basis sets in the Weyl representation (see
e.g.,~\cite{Oz98,Leaf68}): the Heisenberg
 $\hat{T}_\zeta$ basis and
its reciprocal  $\hat{\Delta}_\eta$ basis, where  $\zeta$ and
$\eta$ are  two-component dense indices valued on the standard
symplectic plane of dimension two. Throughout, we denote the
canonical observable pair as $\hat{\eta} = (\hat{x} , \hat{p})$
and use the symplectic product notation $\zeta \wedge \eta \equiv
(\zeta_1 \eta_2 - \zeta_2 \eta_1)$.  The  Heisenberg basis
consists of the translation operators $\hat{T}_\zeta= e^{i
\hat{\eta}\wedge  \zeta  }$ satisfying $ \hat{T}_\zeta^\dagger
\hat{\eta} \hat{T}_\zeta = \hat{\eta} + \zeta \,$, while the
$\hat{\Delta}_\eta$   operators form with the $\hat{T}_\zeta$  a
Fourier-transform pair
\begin{equation}
\hat{\Delta}_\eta =\! \int_\zeta\, e^{ i \zeta\wedge\eta}
\hat{T}_\zeta\, , \ \, \ \hat{T}_\zeta =\int_\eta  e^{ i
\eta\wedge\zeta}\hat{\Delta}_\eta \, .
\end{equation}
The integration measure is defined as $\int_\zeta \equiv
(2\pi)^{-1}\int d^2 \zeta$, etc., with the $2 \pi$ replaced by
$h=2\pi \hbar$ when using units;  note the orthonormality of both
bases with respect to this measure, i.e., $(\eta|\eta') = 2\pi
\delta(\eta-\eta')$, etc. A summary of useful algebraic relations
is
\begin{eqnarray}
\!\!\!\!\hat{T}_\zeta\hat{T}_{\zeta'} =  e^{i \phi_1}
\hat{T}_{\zeta + \zeta'}\, , \  & \
\hat{\Delta}_\eta\hat{\Delta}_{\eta'} =  4 e^{ i
\phi_2}\hat{T}_{2(\eta\!-\!\eta')} \, , \\
\!\!\!\!\hat{T} _\zeta \hat{\Delta} _\eta  = e^{i \phi_3 }
\hat{\Delta}_{\eta+\zeta/2}\, , \ & \ \hat{\Delta}_\eta\hat{T}
_\zeta
  = e^{i \phi_3 } \hat{\Delta}_{\eta-\zeta/2}\, ,
\end{eqnarray}
where $\phi_1 =\frac{1}{2} \zeta' \wedge \zeta $, $\phi_2 =2 \eta
\wedge \eta'$ and $\phi_3 = \eta\wedge \zeta$. For the expansion
of an observable  $\hat{A}$ in either basis
\begin{equation}
\hat{A} = \int_\zeta\, a(\zeta)\, \hat{T}_\zeta = \int_\eta\,
\widetilde{a}(\eta)\, \hat{\Delta}_\eta\, ,
\end{equation}
the expansion functions $a(\zeta)$, and $\widetilde{a}(\eta)$ are
respectively known as the Weyl transform and the Weyl symbol of
 $\hat{A}$ and form a Fourier transform pair;
for the  identity operator,  $I(\zeta) = 2\pi \delta(\zeta)$ and
$\widetilde{I}(\eta) = 1$. The exception again is  the weak value
operator $\hat{W}$, which is expanded in the Hermitian conjugate
of the Heisenberg basis $\hat{W} = \int_\zeta w(\zeta)
\hat{T}_\zeta^\dagger$, in which case the Weyl transform of
$\widetilde{w}(\eta)$ corresponds to the function $w(-\zeta)$.
With these conventions, $ W:\hat{A}
 \rightarrow
  \int_\zeta
 a(\zeta) {w}(\zeta) = \int_\eta
 \widetilde{a}(\eta) {\widetilde{w}}(\eta)$.

In the infinite-dimensional case, the  weak value
 assignment associated with the inference problem  involves integral equations
 with solutions obtained by inverting potentially complicated kernels. For the moment, it
 will suffice to show a generic solution for a reduced set of
observable pairs, leaving open the question of the general set of
pairs for which the problem is solvable. We  consider pairs of the
form ($\hat{p}$,$f(\hat{x})$) for functions $f(x)$ satisfying
conditions determined by the generic solution. Letting
$\hat{U}(t_1,t_2)=\exp[ i(f(\hat{x}) t_2 -\hat{p}\, t_1)]$, the
problem is then to find $w_{\kappa \phi}(\zeta)$ such that for all
$(t_1,t_2)$,
\begin{equation}
e^{i (\phi t_2-\kappa t_1) } = \int_\zeta K(t|\zeta)\, w_{\kappa
\phi}(\zeta)\, ,
\end{equation}
where the kernel $K(t|\zeta) = {\rm Tr}\left[\hat{U}(t_1,t_2)
\hat{T}_\zeta^\dagger\right]$ can be shown to take the form
\begin{equation}
K(t|\zeta) = \delta(\zeta_1-t_1)\int_{-\infty}^{\infty}\!\! du\,
e^{i \left[ \, g(u|\zeta_1)\, t_2 -  \zeta_2\,  u \right] }\, ,
\end{equation}
with $ g(u|\zeta_1)=\frac{1}{2}\int_{-1}^{1}\!ds\,f(u-s
\zeta_1/2)\, $. If $f(x)$ allows the exchange  $\int d\zeta_2 \int
du \leftrightarrow \int du \int d\zeta_2$ (see~\cite{AlbMazz04}),
then it is easily verified that $w_{\kappa \phi}(\zeta)$ must be
the Fourier transform with respect to the $\zeta_2$ variable of a
$\delta$-function at some real root $u_{\phi}(\zeta_1)$ of the
equation $g(u|\zeta_1)=\phi$. Thus,
\begin{equation}
w_{\kappa \phi}(\zeta) = e^{  i\left(u_{\phi}(\zeta_1)
\zeta_2-\kappa \zeta_1\right)}\, .
\end{equation} The function   $f$ must therefore  be such that
$g(u|\zeta_1)=\phi$  admits a real branch for all $\zeta_1$ for
given $\phi$. These conditions admit solutions for polynomial
$f(x)$, but of odd degree only. We have  yet to find solutions for
bounded $f(x)$ (one side or both).

 Proceeding with the realization
of solution ensembles, it proves instructive to revisit the
$(\hat{x},\hat{p})$ inference problem from a constructive
viewpoint. From the definition of $w(\zeta)$ or equivalently from
the above results for $f(x)=x$, we have
 $W: \hat{T}_\zeta \rightarrow w(\zeta) = e^{i \eta_s \wedge \zeta
 }$ for some $\eta_s = (x,p)$. When the problem is stated
in the reciprocal basis, this amounts to finding conditions for
which $\widetilde{w}(\eta) = 2\pi\delta(\eta - \eta_s)$, or
equivalently, $ \hat{W} = \hat{\Delta}_{\eta_s}$. We can now work
in analogy with the discrete case by introducing a maximally
entangled state $\ket{\Phi_0} \equiv \int dx \ket{x}_s\ket{x}_a$,
and two derived basis sets for the combined Hilbert space:
\begin{equation}
 \ket{\Phi_\zeta} \equiv \hat{T}_\zeta \ket{\Phi_0} \, \ \ \ {\rm and}\ \ \ \ket{\Psi_\eta}
\equiv \hat{\Delta}_\eta \ket{\Phi_0}\, ,
\end{equation}
with $\amp{\Psi_\eta}{\Phi_\zeta} =e^{i \eta\wedge\zeta  }$.
Tracing out the ancilla from the outer product
$\ket{\Phi_{\zeta_i}}\bra{\Psi_{\eta_f}}$ for two specific states
and normalizing, we find that
\begin{equation}
\hat{W} = e^{i \zeta_f\wedge \eta_i } \hat{T}_{\zeta_i}
\hat{\Delta}_{\eta_i} = \hat{\Delta}_{\eta_f + \zeta_i/2}\, ,
\end{equation}
which is the desired result with $\eta_s = \eta_f +\zeta_i/2$. One
can verify that   $\ket{\Phi_\zeta}$, $\ket{\Psi_\eta}$ are indeed
$\ket{\xmin,\pplus}$, $\ket{\xplus,\pmin}$ with
$\zeta=(\xmin,\pplus)$, $\eta=(\xplus,\pmin)$.

 One noteworthy
aspect of this solution is the delta-function Weyl symbol for
$\hat{W}$. The ensemble therefore has the remarkable property that
the weak value of any observable is the respective Weyl symbol
evaluated at $\eta=\eta_s$. Thus, when observables with classical
Weyl symbols are weakly (hence, unsharply) probed, the conditional
effects on the instruments from this ensemble will be
indistinguishable from those of a classical system with definite
phase-space localization. (see also~\cite{AB05}).

Any semblance of standard textbook classicality proves illusory,
of course, in  light of the full repertoire of nonlocal
observables  that could be probed weakly. Adapting
equation~(\ref{Krausfour})  to the present language by introducing
the time inversion operation $\zeta^T =(\zeta_1,-\zeta_2)$, we
find for the composite system the $\hat{W}^{(s,a)}$ Weyl transform
 $w^{(s,a)}(\zeta,\zeta') = e^{i (
\eta_s\wedge\zeta + \eta_a\wedge\zeta'
)}e^{\frac{i}{2}\zeta\wedge(\zeta')^T}$ where for the ancilla,
$\eta_a = (x_a,p_a) =(\eta_f - \zeta_f/2)^T$. Note that time
inversion arises naturally from $\hat{T}_\zeta^{(a)}\ket{\Phi_0} =
T_{\zeta^T}^{(s)}\ket{\Phi_0}$. For  test functions,  the
corresponding Weyl symbol is then $
\widetilde{w}^{(s,a)}(\eta,\eta') = e^{-\frac{i}{2} \partial_\eta
\wedge
\partial_{\eta'}^T }\delta(\eta-\eta_s)\delta(\eta'-\eta_a) \, .
$
Thus, while joint weak measurements yield Weyl symbols for system
and ancilla observables $\hat{A}$ and $\hat{B}$ respectively, weak
measurements of $\hat{A}\hat{B}$ will differ from the product of
the Weyl symbols by terms involving ``crossed" derivatives to all
orders (equivalently, the weak value of $\hat{A}\hat{B}$
determines correlations in the outcomes of the joint weak
measurement of $\hat{A}$ and $\hat{B}$~\cite{ReStei04}). The
signature of entanglement is therefore a joint Weyl symbol that is
essentially a non-local object in the combined ``phase-space".

We conclude with some remarks on interpretational issues. First,
it is hard to ignore the fact that for those conditions solving
the canonical inference problem, the conditionally sharp outcomes
$(x,p)$ are in consistent correspondence with  numbers
representing physical properties of the system in other contexts.
To wit: we find algebraic correspondence with the initial and
final state labels through the  transformations of
Eq.~(\ref{transeq}), and at the level of weak values, with
phase-space functions evaluated at $(x,p)$.
 These correspondences seem to indicate a certain inner consistency of
  the quantum framework which, under special circumstances,
allows for a statistically unambiguous operational assignment of
values to both canonical variables. Finally, it has been suggested
earlier\cite{AV90,AV02} that  it is not one, but rather a pair of
state vectors (e.g., our initial and final states) that provide
the complete description of a quantum system at a given instant.
The present results  provide a good indication of the extent to
which the descriptive power of the two-vector framework is
enhanced by the property of entanglement. It appears that through
entanglement, the framework becomes flexible enough to
incorporate, among many other possibilities, the classical
description of a single system, at least at the kinematic level.
We believe this realization raises interesting questions,
particularly regrading our perception of classicality in the
macroscopic domain.

The author is grateful to P. Mazur and J. Suzuki for helpful
discussions.

\end{document}